%% file: main.tex
\documentclass[10pt,a4paper, fleqn]{article}

\input{style/packages}
\input{style/commands}

\author{Lea F\"ocke$^{1,2}$}
\newcommand{\simrx}{\emph{SiMRX} }
\newcommand{\version}{1.4}
\title{\simrx -- A \emph{Si}mulation toolbox for \emph{MRX}\\{\footnotesize Technical Report for \simrx version \version}}

\begin{document}

\maketitle
\input{sections/abstract}
\input{sections/introduction}
\input{sections/structure}
\input{sections/features}

\input{sections/setups}

\input{sections/examples}
\input{sections/remarks}

\nocite{*}%
\printbibliography%

\appendix
\input{sections/documentation}

\end{document}

%% file: style/packages.tex
\usepackage{authblk}

\usepackage[utf8]{inputenc}
\usepackage[backend=biber,%
style=numeric-comp,%
isbn=false,%
sorting=nyt,%
doi=false,%
url=false,%
backref,%
backrefstyle=three]{biblatex}
\usepackage{csquotes}
\usepackage[english]{babel}

\usepackage{blindtext}
\usepackage{microtype}
\usepackage{float}
\usepackage{paralist}
\usepackage{pmboxdraw}

\usepackage{abstract}

\providecommand{\keywords}[1]{\textbf{\textit{keywords---}} #1}

\usepackage{amsmath}
\usepackage{amssymb}
\usepackage{amstext}
\usepackage{amsfonts}
\usepackage{amsthm}

\usepackage{setspace}
\usepackage{array}
\usepackage[export]{adjustbox}
\usepackage{graphicx}
\usepackage[table]{xcolor}

\usepackage{subcaption}
\usepackage{dsfont}
\usepackage{shadethm}
\usepackage{wasysym}
\usepackage{listings}
\usepackage{algorithm}
\usepackage{algpseudocode}
\usepackage{parskip}
\usepackage[bookmarks,raiselinks,pageanchor,colorlinks,%
citecolor=black,linkcolor=black,urlcolor=black,filecolor=black,menucolor=black]{hyperref}
\usepackage[noabbrev]{cleveref}
\usepackage{contour}
\usepackage{accents}
\usepackage{multirow,tabularx}
\newcolumntype{B}[1]{>{\centering\arraybackslash}m{#1}}
\newcolumntype{L}[1]{>{\arraybackslash}m{#1}}
\newcolumntype{R}[1]{>{\raggedleft\arraybackslash}m{#1}}

\usepackage{nomencl}
\usepackage{capt-of}
\usepackage{afterpage}
\usepackage{ltxtable}  
\usepackage{enumitem}
\usepackage{stmaryrd}
\usepackage{siunitx}

\usepackage{todonotes}
\usepackage{a4wide}

\usepackage[numbered]{matlab-prettifier}

\lstdefinestyle{matstyle}{%
	style              	= Matlab-editor,
	basicstyle			=\scriptsize\ttfamily,
	escapechar 			= ",
	mlshowsectionrules 	= true,
}
\lstdefinestyle{pystyle}{%
	style              	= Matlab-Pyglike,
	basicstyle			=\scriptsize\ttfamily,
	escapechar 			= ",
	mlshowsectionrules 	= false,
}

\usepackage[format=plain,singlelinecheck=false, margin={\leftmargin, 0pt}]{caption}

\newfloat{lstfloat}{t}{lop}
\floatname{lstfloat}{Listing}

\crefname{lstfloat}{listing}{listings}
\Crefname{lstfloat}{Listing}{Listings}

\newlength{\mytextwidth}
\newenvironment{mymarginbox}[1]{\hspace*{\leftmargin}\begin{minipage}{\mytextwidth}#1}{\end{minipage}}

\makeatletter
\renewcommand{\maketitle}{\par
\begingroup
	\renewcommand\thefootnote{\@fnsymbol\c@footnote}%
	\def\@makefnmark{\rlap{\@textsuperscript{\normalfont\@thefnmark}}}%
	\long\def\@makefntext##1{\parindent 1em\noindent
	      \hb@xt@1.8em{\hss\@textsuperscript{\normalfont\@thefnmark}}##1}%
	\if@twocolumn
	\ifnum \col@number=\@ne
	  \@maketitle
	\else
	  \twocolumn[\@maketitle]%
	\fi
	\else
	\newpage
	\global\@topnum\z@   
	\@maketitle
	\fi
	\thispagestyle{plain}\@thanks
\endgroup
\setcounter{footnote}{0}%
\global\let\thanks\relax
\global\let\maketitle\relax
\global\let\@maketitle\relax
\global\let\@thanks\@empty
\global\let\@author\@empty
\global\let\@date\@empty
\global\let\@title\@empty
\global\let\title\relax
\global\let\author\relax
\global\let\date\relax
\global\let\and\relax
}
\def\@maketitle{%
  \newpage
  \null
  \begin{mymarginbox}%
  \let \footnote \thanks
    {\LARGE \@title \par}%
    \vskip 1.5em%
    {\large \@author\par}%
    \vskip 1em%
    {\large \@date}%
  \end{mymarginbox}%
  \par
  \vskip 1.5em}
\makeatother

%% file: style/commands.tex
\newcommand{\R}{\mathbb{R}}
\newcommand{\N}{\mathbb{N}}

\newcommand{\C}{\mathcal{C}}

\newcommand\cellwidth{\TX@col@width}

\newcommand{\mypath}[1]{%
\begingroup\setlength{\fboxsep}{1pt}%
\colorbox{lightgray!30}{\texttt{\hspace*{2pt}\vphantom{Ay}#1\hspace*{2pt}}}
\endgroup}

%% file: sections/abstract.tex
\begin{abstract}
\simrx is a MRX simulation toolbox written in MATLAB for simulation of realistic 2D and 3D \textit{Magnetorelaxometry} (MRX) setups, including coils, sensors and current patterns.
MRX is a new modality that uses magnetic nanoparticles (MNP) as contrast agent and shows promising results in medical applications, e.g. cancer treatment.
Its basic principles were outlined in \cite{baumgarten2008MagneticNanoparticleImaging}, further elaborated in \cite{liebl2014QuantitativeImagingMagnetic}, transferred into a rigorous mathematical model and analyzed in \cite{focke2018InverseProblemMagnetorelaxometry}.

\simrx is available at \url{https://gitlab.com/simrx/simrx/}.
\end{abstract}

\footnotetext[1]{Institute for Analysis and Numerics, Westfälische Wilhelms-Universität Münster (WWU), DE}
\footnotetext[2]{Applied Mathematics, Friedrich-Alexander Universität Erlangen-Nürnberg (FAU), DE}
\footnotetext[3]{Institute of Biomedical Engineering and Informatics, Technische Universität Ilmenau (TU Ilmenau), DE}
\footnotetext[4]{Institute of Electrical and Biomedical Engineering, Private University of Health Sciences, Medical Informatics and Technology (UMIT), Hall in Tirol, AT}
\footnotetext[5]{Physikalisch-Technische Bundesanstalt (PTB), Berlin, DE}
\vspace{0.3cm}

\renewcommand{\abstractname}{Acknowledgements and Contributions}
\begin{abstract}
This toolbox was inspired by earlier work of Peter Hoemmen\footnotemark[3], Paul Koenigsberger\footnotemark[3] and Daniel Baumgarten\footnotemark[3]$^,$\footnotemark[4].
In particular for the simulation of the 3D case, we recycled code fragments originally created by Peter Hoemmen\footnotemark[3] and Paul Koenigsberger\footnotemark[3].
Moreover, the 3D simulation step has been validated using a measured 3D dataset provided by Maik Liebl\footnotemark[5].
The phantoms were created as part of a collaboration with Daniel Baumgarten\footnotemark[3]$^,$\footnotemark[4] and Peter Schier\footnotemark[4].
As part of the ongoing collaboration with Daniel Baumgarten\footnotemark[3]$^,$\footnotemark[4] and Peter Schier\footnotemark[4] this code may receive updates and new features in the future.

Maik Liebl has been supported by the German Science Foundation (DFG) within the priority program SPP1681 (WI 4230/1-3).

This work has been supported by the German Science Foundation (DFG) within the priority program CoSIP, project CoS-MRXI (BA 4858/2-1, BU 2327/4-1).
\end{abstract}
\keywords{{\footnotesize magnetorelaxometry imaging; magnetic nanoparticles; modeling; MATLAB toolbox; \simrx}}

%% file: sections/introduction.tex
\section{Introduction}
Many new and experimental treatment methods in medical applications use magnetic nanoparticles as a contrast agent.
These particles allow for multiple different approaches (\cite{hiergeist1999ApplicationMagnetiteFerrofluids, alexiou2011CancerTherapyDrug}), however for the named methods the exact knowledge about the particle distribution is crucial.
Here, Magnetorelaxometry (MRX) can be used to determine the amount of particles in a region of interest as shown in \cite{baumgarten2008MagneticNanoparticleImaging, liebl2014QuantitativeImagingMagnetic}.
Based on this approach an imaging technique called Magnetorelaxometry Imaging (MRXI) has been proposed.

The \simrx toolbox provides a set of tools to model and simulate such an MRX system.
It is based on the mathematical model developed in \cite{focke2018InverseProblemMagnetorelaxometry}.

\subsection{Model}
This following section (including notation) is part of \cite{focke2018InverseProblemMagnetorelaxometry}, which, for interested readers, provides an in depth look in the analysis of the following operator.
The magnetic field in $w\in\Omega$ induced by a coil $\alpha = (\varphi_\alpha, I_\alpha)$ is given by
\begin{equation}
\mathbf{B}_\alpha^\textbf{coil}\colon\Omega \rightarrow\R^3, \quad w \mapsto\vartheta\int\limits_0^{L_\alpha}\varphi_\alpha^\prime(s)\times\left(\frac{w-\varphi_\alpha(s)}{\left\vert w-\varphi_\alpha(s)\right\vert^3}\right)ds,\label{eq:biotsavart}
\end{equation}
where ${L_\alpha}$ is the length of the coil.
The magnetization of the magnetic nanoparticles (MNP) after the reorientation process in $w$ is described by
\begin{equation}
\mathbf{m}_\alpha\colon \Omega\rightarrow\R^3, \quad w \mapsto\frac{1}3 \mathbf{B}_\alpha^\textbf{coil}(w)c(w),\label{eq:dipolemagnetization}
\end{equation}
where $c$ is the desired particle distribution.
The particle induced magnetic response from particles in $w$ measured by a sensor $\sigma = (\sigma_x, \sigma_n)$ is modeled by
\begin{align}
	\mathbf{B}^\textbf{meas}_\alpha\colon\Omega\times\Sigma&\rightarrow\R\nonumber\\
	\left(w,\sigma\right)&\mapsto\sigma_n\cdot\left(\left(\frac{3\left(\sigma_x-w\right)\otimes\left(\sigma_x-w\right)}{\left\vert\sigma_x-w\right\vert^5}-\frac{\mathbb{I}}{\left\vert\sigma_x-w\right\vert^3}\right)\mathbf{B}_\alpha^\textbf{coil}(w)\right).\label{eq:measmagnetization}
\end{align}
In the end we receive the following forward operator for a coil $\alpha$:
\begin{equation}
\mathbf{K}_\alpha\colon\mathcal{L}^2(\Omega) \rightarrow\mathcal{L}^2(\Sigma), \quad c \mapsto\left[\sigma\mapsto\int\limits_\Omega\mathbf{B}^\textbf{meas}_\alpha(w, \sigma)\ c(w) d^3w\right].\label{eq:fredholmformulation}
\end{equation}

\subsection{Discretization}\label{ss:coilDiscretization}
First we consider the 3D case:
Here the conductor coil $\varphi_\alpha$ is approximated by a set list of segments, with starting points $a_k$ and ending points $b_k$ for the $k$-th segment respectively.
Then the magnetic field in $w\in\Omega$ is \cite{hanson2002CompactExpressionsBiot}:
\begin{align}
	\mathbf{B}_{\alpha,k}^\textbf{coil}\colon\Omega&\rightarrow\R^3\label{eq:discreteActivation}\nonumber\\
	w&\mapsto\vartheta\frac{\vert a_k-w\vert+\vert b_k-w\vert}{\vert a_k-w\vert\vert b_k-w\vert}\frac{(a_k-w)\times(b_k-w)}{\vert a_k-w\vert\vert b_k-w\vert+(a_k-w)\cdot(b_k-w)}.
\end{align}
The MNP response (see equation \eqref{eq:measmagnetization}) still holds in the discrete case.

In the 2D case a coil simplification leads to the usage of variants of \eqref{eq:measmagnetization} for both coil and dipole response (see \cite[section 4.2]{focke2018InverseProblemMagnetorelaxometry}).
This is based on the idea tha -- from a certain distance -- a coil induced magnetic field is structurally indistinguishable to a appropriately strong dipole field.

%% file: sections/structure.tex
\section{Structure}
The \simrx toolbox is a modular toolkit that provides tools for MRX experiment setup configuration and simulation as well as visualization of data.
\simrx is capable of simulating synthetic or real setups and datasets.

The processes required for simulation are handled in a sequence of modular functions, i.e. the creation of voxel grids, the calculation of magnetic fields and ultimately the data acquisition at the sensors.
See \cref{ssec:simulation} for a detailed description of the simulation step and relate to \cref{im:systemmatrixPattern} and \cref{im:systemmatrixCoil} for an illustration of the simulation process.

For the simulation of an MRX experiment we introduce a distinction between a \texttt{setup} and \texttt{config} file.
The \texttt{setup} file includes information about the number of dimensions, coil position and shape (and, if necessary, orientation), sensor position and orientation, as well as intervals that define the region of interest.
The \texttt{config} file includes information about the resolution of the phantom within the region of interest, coil current patterns, as well as information what subset of coils and sensors are actively used.
Both files datasets are forwarded to the simulation script, that return a matrix representation of the given setup and configuration.
Detailed information on the creation of \texttt{setup} and \texttt{config} file can be found in \cref{ss:config}.

Furthermore \simrx provides useful tools that visualize MRX setups and internal states (\cref{ss:vis}), and includes a tool for the creation of phantoms (\cref{ss:phantom}).

\textbf{Please note:} The \simrx toolbox used a unified coordinate system, where spatial information is stored in a three dimensional vector $(x,y,z)$.
Any position is given in relation to a zeropoint $(0,0,0)$ and, if not specified further, is provided in $\left[\si{m}\right]$.

If \simrx is used to implement a 2D setup, please read the paragraph about \textit{Conventions for 2D configurations} in the introduction of \cref{para:config}.

%% file: sections/features.tex
\section{Features and Modules}\label{sec:features}
The \simrx toolbox is separated in the following submodules (each in its own subfolder):

\begin{mymarginbox}	
	\begin{tabular}{@{}L{\textwidth}@{}}%
		\mypath{./configuration}\\
		\mypath{./simulation}\\
		\mypath{./phantom}\\
		\mypath{./visualization}
	\end{tabular}
\end{mymarginbox}

We give a short overview of each module in the following sections.
For all provided functions, in detail information of syntax and features are available in each file header.
This documentation is also available using the MATLAB help function (F1).

\subsection{Configuration}\label{ss:config}
\begin{figure}
	\begin{mymarginbox}
		\begin{lstfloat}[H]
\begin{lstlisting}[style=matstyle]
%% create a new 2D setup
setup               = [];
setup.info.name     = '2Dsetup';

setup.dim           = 2;

setup.roi.x         = [-.05, .05];
setup.roi.y         = [-.05, .05];
setup.roi.z         = [   0,   0];

coils               = [];
coils = [coils, createEntityArray([ .055,.045,0], [ .055,-.045,0], [-1,0,0], [1,10,1])];
coils = [coils, createEntityArray([-.055,.045,0], [-.055,-.045,0], [ 1,0,0], [1,10,1])];
coils = [coils, createEntityArray([.045, .055,0], [-.045, .055,0], [0,-1,0], [10,1,1])];
coils = [coils, createEntityArray([.045,-.055,0], [-.045,-.055,0], [0, 1,0], [10,1,1])];
setup.coils         = coils;
setup.coilGroups    = {1:10, 11:20, 21:30, 31:40};

sensors             = [];
sensors = [sensors, createEntityArray([.050,.060,0], [-.050,.060,0], [-1,-1,0], 10)];
sensors = [sensors, createEntityArray([.050,.060,0], [-.050,.060,0], [ 1,-1,0], 10)];
setup.sensors       = sensors;
setup.sensorGroups  = {1:10, 11:20};

visualizeMRX(setup);

%% validate and save setup
setup.info.variant  = 'default';
if isSetupValid(setup)
    saveSetup(setup.info.variant, setup);
end

\end{lstlisting}
			\caption[2D setup]{MATLAB code to create a 2D setup called \emph{2Dsetup}. The region of interest is defined as a 2D square with a side length of 10\si{cm}. Ten activation coils are positioned in front of each the edge of the region of interest. Each sides coils are collected into a unique coil group, which lead to four coil groups. A total of 20 sensors are positioned above the region of interest and have an angled orientation. The different orientations are grouped accordingly as sensor groups. Since no alternations to the coils are done, the 2D default model of the coils is used, hence \emph{default} is chosen as variant name.}
			\label{code:2Dsetup}
		\end{lstfloat}
	\end{mymarginbox}
\end{figure}
\begin{figure}
	\begin{mymarginbox}
		\begin{lstfloat}[H]
\begin{lstlisting}[style=matstyle]
%% create a new 2D config
config                      = [];

config.info.name            = 'all';
config.coilsActive          = unique([setup.coilGroups{:}]);
config.sensorsActive        = unique([setup.sensorGroups{:}]);

config.info.variant         = 'singleSequential';
config.currentPattern       = createPattern(config.coilsActive, 'sequential', ...
                                'current', 0.8);
config.measurementPattern   = createPattern(config.sensorsActive, 'simultaneous', ...
                                'times', size(config.currentPattern, 1));

%% check compatibility with setup and save config
if checkCompatibility(setup, config)
    saveConfig(sprintf('configs/%s/%s', config.info.name, config.info.variant), config);
end

\end{lstlisting}
			\caption[2D config]{MATLAB code to create a config suitable for the setup created in \cref{code:2Dsetup}. The configuration name is set to \emph{all} since it uses all coils and sensors of the corresponding setup. Then a subsequent current pattern, which only one coil active at a time is implemented. Here all sensors are active for every activation. This config variant is named \emph{singleSequential}.}
			\label{code:2Dconfig}
		\end{lstfloat}
	\end{mymarginbox}
\end{figure}
This module provides a set of functions to create \texttt{setup} and \texttt{config} files.
\paragraph{Conventions for 2D Configurations:}\label{para:config}
\simrx supports 2D and 3D systems.
However, at its core, the toolbox is designed for 3D simulations.
Therefore 2D setups are implemented as 3D setups with the following adjustments:
\begin{itemize}
\item all entities are on the same z-layer (z=0 recommended)
\item the region of interest in z direction is set accordingly (\texttt{setup.roi.z = [0,0]} recommended)
\item the voxel resolution \texttt{res} in z direction is set to 1, e.g. \texttt{res = [x,y,1]};
\item coils \emph{must be} represented as dipole magnets to be in line with \cref{ss:coilDiscretization}. This means that \texttt{coils.Segments} will be ignored, regardless of availability.
\end{itemize}
Also consider the provided 2D and 3D setups as part of \simrx (see \cref{subs:sytheticdataset}).

\subsubsection{Setup file}\label{sss:setup}
In MATLAB a \texttt{setup} file is represented as struct.
It holds the following fields

\begin{mymarginbox}%
	\begin{tabular}{l}%
		\texttt{info},\quad\texttt{dim},\quad\texttt{roi},\quad\texttt{coils},\quad\texttt{coilGroups},\quad\texttt{sensors}\quad and\quad\texttt{sensorGroups}.
	\end{tabular}
\end{mymarginbox}

As of version \version\ \texttt{setup.info} struct only contains the name of the setup, as well as the current variant name (for setup variants, see \cref{ss:setupfolderstructure}).

Determined by the dimension of the system, \texttt{setup.dim} is either 2 or 3.
The region of interest \texttt{setup.roi} is a struct with fields \texttt{x}, \texttt{y} and \texttt{z}, that holds boundary information in an array with unit meter $\left[\si{m}\right]$. For instance \texttt{setup.roi.x = [0,0.1]} translates to a region of interest of $10\si{cm}$ length in x direction.
Yet again \texttt{setup.coils} and \texttt{setup.sensors} are arrays of structs with fields:

\begin{mymarginbox}%
	\begin{tabular}{@{}L{0.30\textwidth}@{}L{0.30\textwidth}@{}}%
		\texttt{coils(i).Position},&\texttt{sensors(i).Position},\\
		\texttt{coils(i).Normal},&\texttt{sensors(i).Normal},\\
		\texttt{coils(i).Segments},
	\end{tabular}
\end{mymarginbox}

where $i$ is the $i$-th coil/sensor.
As mentioned before, spatial information (in this case the fields \texttt{Position} and \texttt{Normal}) are given as a three element vector with base unit meter $\left[\si{m}\right]$.

The function \texttt{createEntityArray.m} is used to create new coil and sensor entries.
In the current version the function takes two outer points, a normal and the number of intermediate points.
Then \texttt{createEntityArray.m} returns a list of equidistant points, that represent the entity position, as well as the respective normal.

The setup format allows to group coil and sensors.
By this the creation of config files for certain partitions of a setup can be streamlined.
This is realized by the cell arrays \texttt{setup.coilGroups} and \texttt{setup.sensorGroups}, where each cell element contains a list of coil and sensor IDs respectively.

In the 3D case the coil wire is modeled by a list of points, which represent the coil shape.
The function \texttt{createCoilLoop.m} creates a coil template, which has to be copied in every coil position.
The function \texttt{parseCoils.m} receives a coil struct (with fields \texttt{Position} and \texttt{Normal}) and adds the \texttt{Segments} field in respect to the given coil template.
Internally this uses the function \texttt{relocateStructure.m} that reorientates a given list of points.

In case a predefined voxel grid is given, the corresponding region of interest can be determined using \texttt{getROI.m}. During the creation of a new \texttt{setup} the usage of \texttt{visualizeSetup.m} is recommended.

The script in \cref{code:2Dsetup} can be used to create the \texttt{setup} file for the 2D system shown in \cref{im:vis}.
Please also examine the examples given in \cref{ss:examplesetups}, that make use of the described functionalities.

\subsubsection{Config file}\label{sss:config}
\texttt{config} files are, similar to \texttt{setup} files, structs in MATLAB. They hold the fields

\begin{mymarginbox}
	\begin{tabular}{l}%
		\texttt{info},\quad\texttt{coilsActive},\quad\texttt{sensorsActive},\quad\texttt{currentPattern}\quad and\ \texttt{measurementPattern}.
	\end{tabular}
\end{mymarginbox}

As for the \texttt{setup} struct, the \texttt{config.info} struct only contains the name of the config, as well as the current variant name (for config variants, see \cref{ss:setupvariants}).

A setup may contain a lot of coils/sensors that are not used in a specific configuration.
To optimize the simulation speed \texttt{coilsActive} and \texttt{sensorsActive} are lists of coil and sensor ids, that define the subset of active setup parts.

The coil current sequence is stored in \texttt{config.currentPattern} as a matrix, where each column corresponds to a coil.
Each row represents one pattern, where the entries refer to the applied current at the respective coil.
The measurement pattern is stored in \texttt{config.measurementPattern} in binary matrix form.
Here each column corresponds to a sensor, which is active during the current measurement depending on the binary entry.
The function \texttt{createPattern.m} provides standard patterns and multiple other presets and can be used to even-handedly create current and measurement patterns.

The script in \cref{code:2Dconfig} can be used to create a \texttt{config} file for the 2D system shown in \cref{im:vis}.
Again please examine the examples given in \cref{ss:examplesetups}, that make use of the described functionalities.

\subsubsection{Setup and Config Validation}\label{sss:scvalid}
This toolbox also contain functions to check the created \texttt{setup} and \texttt{config}.
The functions \texttt{isConfigValid.m} and \texttt{isSetupValid.m} validate if all required fields are set and further include checks for data inconsistencies.
Additionally the function \texttt{checkCompatibility.m} checks if a given \texttt{config} is applicable for a given \texttt{setup}.

\subsubsection{Save and Load}
The \texttt{setup} and \texttt{config} datasets can be stored using \texttt{saveSetup.m} and \texttt{saveConfig.m}.
These are stored using a custom \texttt{.mrxsetup} and \texttt{.mrxcfg} extension, that is based on the MATLAB internal \texttt{.mat} file format.
Corresponding load functions (\texttt{loadSetup.m} and \texttt{loadConfig.m}) are available.

To store the created \texttt{setup} and \texttt{config} datasets, we recommend using the folder structure proposed in \cref{ss:setupfolderstructure}.

\subsection{Simulation}
\label{ssec:simulation}
\begin{figure}[t]
	\begin{mymarginbox}
	\includegraphics[width=\textwidth, page=2]{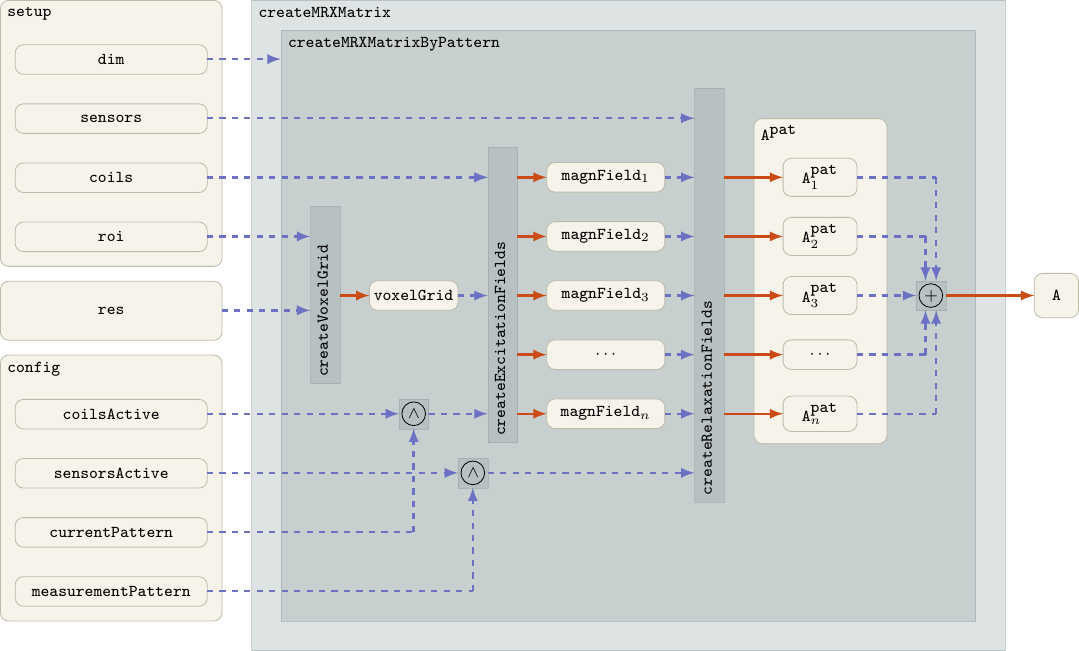}
	\end{mymarginbox}
	\caption{Schematics of the pattern-based system matrix creation workflow. Variables are marked with a rounded corner box, functions are marked with a gray box. The variable usage is color coded: green (dashed) as input and red (solid) as output argument.}
	\label{im:systemmatrixPattern}
\end{figure}
\begin{figure}[t]
	\begin{mymarginbox}
	\includegraphics[width=\textwidth, page=1]{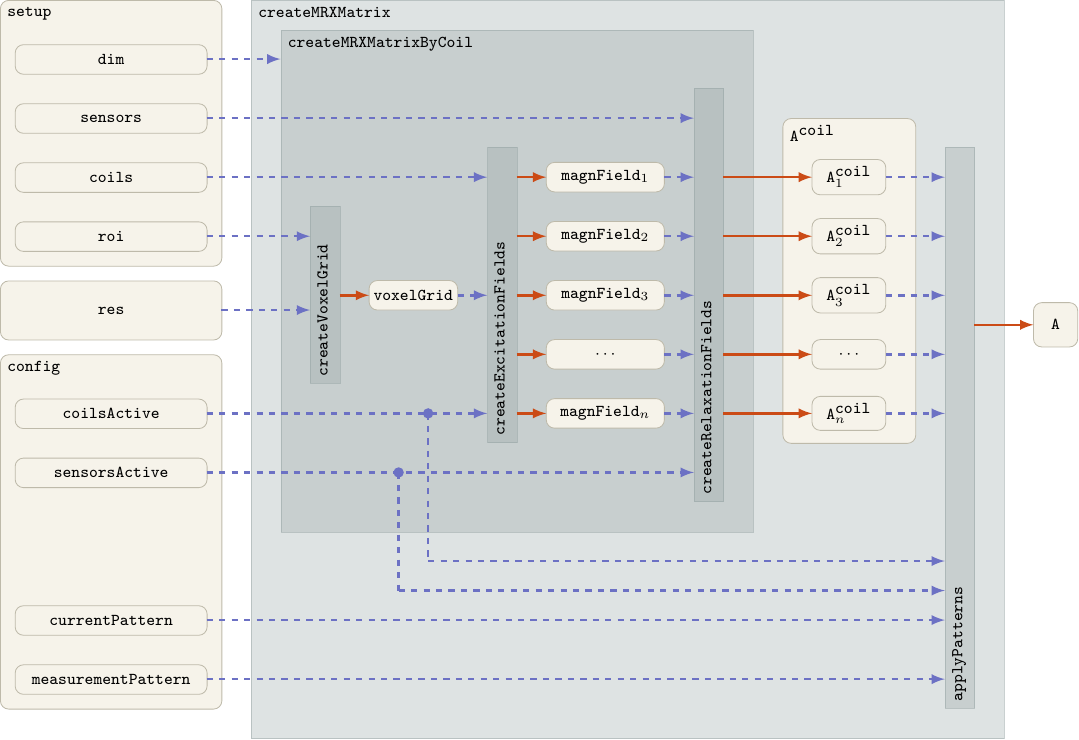}
	\end{mymarginbox}
	\caption{Schematics of the coil-based system matrix creation workflow using raw files internally. Variables are marked with a rounded corner box, functions are marked with a gray box. The variable usage is color coded: green (dashed) as input and red (solid) as output argument.}
	\label{im:systemmatrixCoil}
\end{figure}
The simulation module is the core element of the toolbox.
All necessary files can be found in the folder \mypath{./simulation}.
The main script is provided by the file \texttt{createSystemMatrix.m}.
It processes a given MRX setup and configuration into a linear operator, namely a matrix \texttt{A}.
A valid \texttt{setup} file is required, that provides information about dimension, region of interest, coil position/orientation and sensor position/orientation.
Furthermore lists of active coils and sensors as well as coil current and measurement patterns are provided by a \texttt{config} file.
For the simulation step we require to provide the desired voxel/pixel resolution.
See \cref{sss:scvalid} for tools to validate the setup and config pair.
By default the function \texttt{createSystemMatrix.m} internally calls \texttt{createSystemMatrixByPattern.m}.
Alternatively \texttt{createSystemMatrix.m} can be requested to call the function \texttt{createSystemMatrixByCoil.m} instead.
Both processes create the same system matrix, however it is performed in different ways:
The function \texttt{createSystemMatrixByPattern.m} derives the system matrix one pattern after another.
It combines the active coils magnetic fields with the contemporary coil current on the fly and derives the resulting dipole fields for the final system matrix.
On the other hand \texttt{createSystemMatrixByCoil.m} derives a base magnetic field with a unitary coil current for each coil individually.
Then for each of these fields the resulting dipole fields are calculated and stored as a \emph{raw measurement}.
Due to linearity of the system, the final system matrix can be derived as linear combination of these raw measurements.
These processes are illustrated in \cref{im:systemmatrixPattern} and \cref{im:systemmatrixCoil}.

The internal structure of the \texttt{createSystemMatrixByPattern.m} routine is shown in \cref{im:systemmatrixPattern}.
First a valid \texttt{voxelGrid} is created using \texttt{createVoxelGrid.m}.
Then for each pattern we derive the currently active coils with a activation current according to \texttt{config.currentPattern}.
The compiled list of coils is used in \texttt{createExcitationFields.m} to calculate the resulting magnetic field for this current pattern.
Internally the choice of the used mathematical model depends on the existence of \texttt{coils.Segments} (see \cref{ss:coilDiscretization}).
Finally \texttt{createRelaxationFields.m} translates the derived field on the voxel grid into the magnetic dipole response.
However the evaluation of the magnetic response is only performed on the list of active sensors in respect to \texttt{config.measurementPattern}.
A composition of all patterns lead to the fill system matrix $A$.

\Cref{im:systemmatrixCoil} shows the internal structure of the \texttt{createSystemMatrixByCoil.m} routine.
In the same way \texttt{createVoxelGrid.m} creates a \texttt{voxelGrid} first.
Then \texttt{createExcitationFields.m} is called to calculate the magnetic field for each coil individually with a unitary current.
The base magnetic fields are feed into \texttt{createRelaxationFields.m}.
This function translates the fields present on the voxel grid into the magnetic dipole fields, which is acquired by the sensors.
As a result we receive a \emph{raw measurement} for each coil, which has been created by a single coil with unitary current.
Due to the linearity of the forward operator, we can now apply a current and measurement pattern by using \texttt{applyCurrentPattern.m}.

We see that these procedures have it pros and cons.
The default derivation by pattern is usually faster, since we do not have any overhead in deriving and storing the raw system matrices corresponding to single coil activations.
Depending on the activation and measurement pattern, the function \texttt{createSystemMatrixByCoil.m} may do a lot of unnecessary work.
On the other side however, when for example only small parts of the system are changed between experiments, the function \texttt{createSystemMatrixByPattern.m} recalculates every signal from the ground up, whereas the function \texttt{createSystemMatrixByCoil.m} is able to reuse the raw measurements and only requires to reevaluate the current and measurement patterns.
Given that storage capacity is available, it may often be an advantage to use previously calculated raw files instead, as described in the next section.

\subsubsection{Raw export/import}\label{sss:rawExIm}
As stated in the end of the previous section, it can be advantageous to use precalculated raw files, since the simulation of the setup is a time demanding step in most cases.
The \simrx toolbox provides tools to dump \texttt{ARaw} files into file.
As seen in \cref{im:systemmatrixCoil}, \texttt{ARaw} dataset are preliminary files in the system matrix creation process.
Storing these separately allows for flexible experimentation with varying coil currents.
Note that this procedure is not very suitable in case the set of active coils or sensors changes regularly, since \texttt{ARaw} contains a fixed set of precalculated coil results.

The function \texttt{exportRawSetup.m} allows for in batch export of single coil and sensor data, whereas \texttt{importRawSetup.m} allows to import previously derived raw files.
Internally the simulation steps done in \texttt{exportRawSetup.m} are equivalent to those steps done by \texttt{createSystemMatrixByCoil.m}.
A working example (\texttt{ExampleB.m}) is presented in \cref{s:examples}.

\subsection{Phantom}\label{ss:phantom}
\begin{figure}%
	\begin{mymarginbox}
	\includegraphics[width=\textwidth]{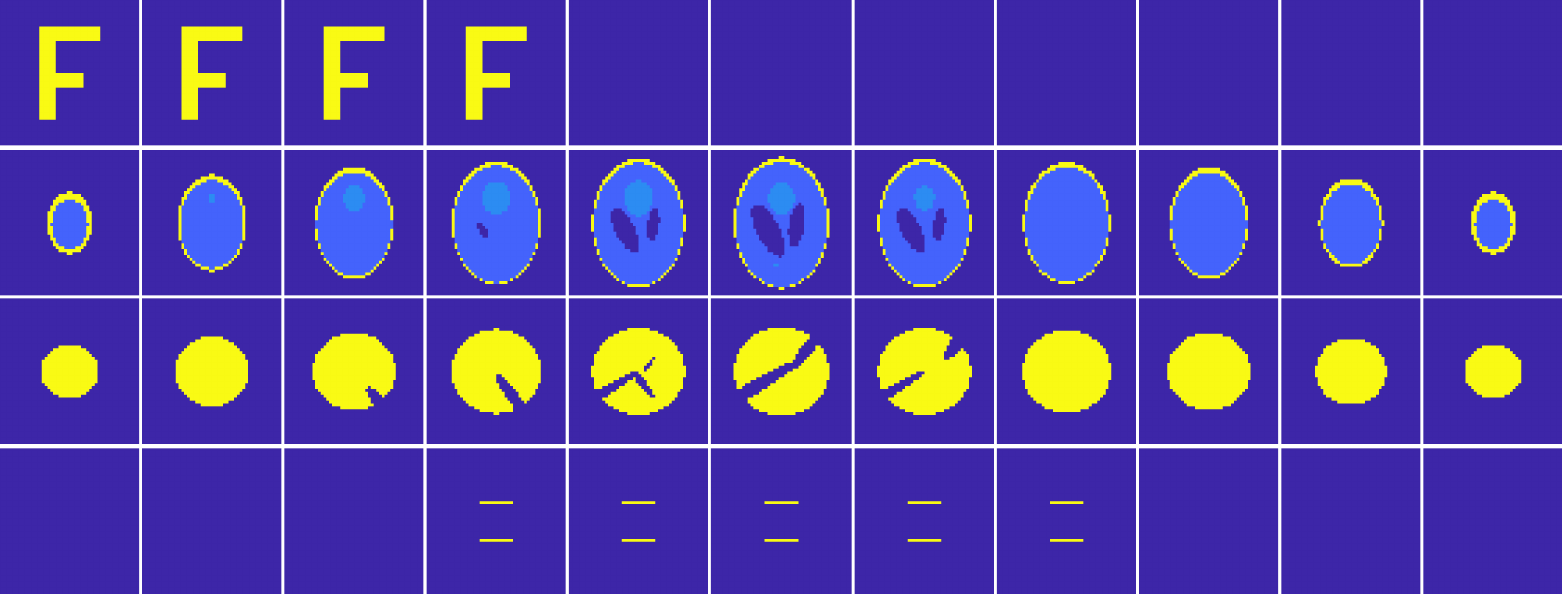}
	\end{mymarginbox}
	\caption{Selection of available 3D phantoms with a resolution of $[50,50,15]$. One phantom per row, each column represents a layer. Layers $1, 2, 14$ and $15$ are cut from illustration, due to clarity. Phantom names from top to bottom: \texttt{'F\_2'},\texttt{'shepplogan3d'},\texttt{'tumor'} and \texttt{'fwhmdots\_0.25'}.}
	\label{im:phantom}
\end{figure}
The function \texttt{createPhantom.m} provides multiple phantom options, including phantoms suited for reconstruction or resolution tests.
\texttt{createPhantom.m} uses the function \texttt{phantom3.m}, which is a 3D reimplementation of the MATLAB given phantom function and able to generate 3D phantoms that are composed of multiple ellipses.
For all available phantoms, please check the help section of \texttt{createPhantom.m} and \texttt{phantom3.m}. See \cref{im:phantom} for a selection of the available phantoms.

\subsection{Visualization}\label{ss:vis}
\begin{figure}
	\begin{mymarginbox}
		\includegraphics[width=\textwidth]{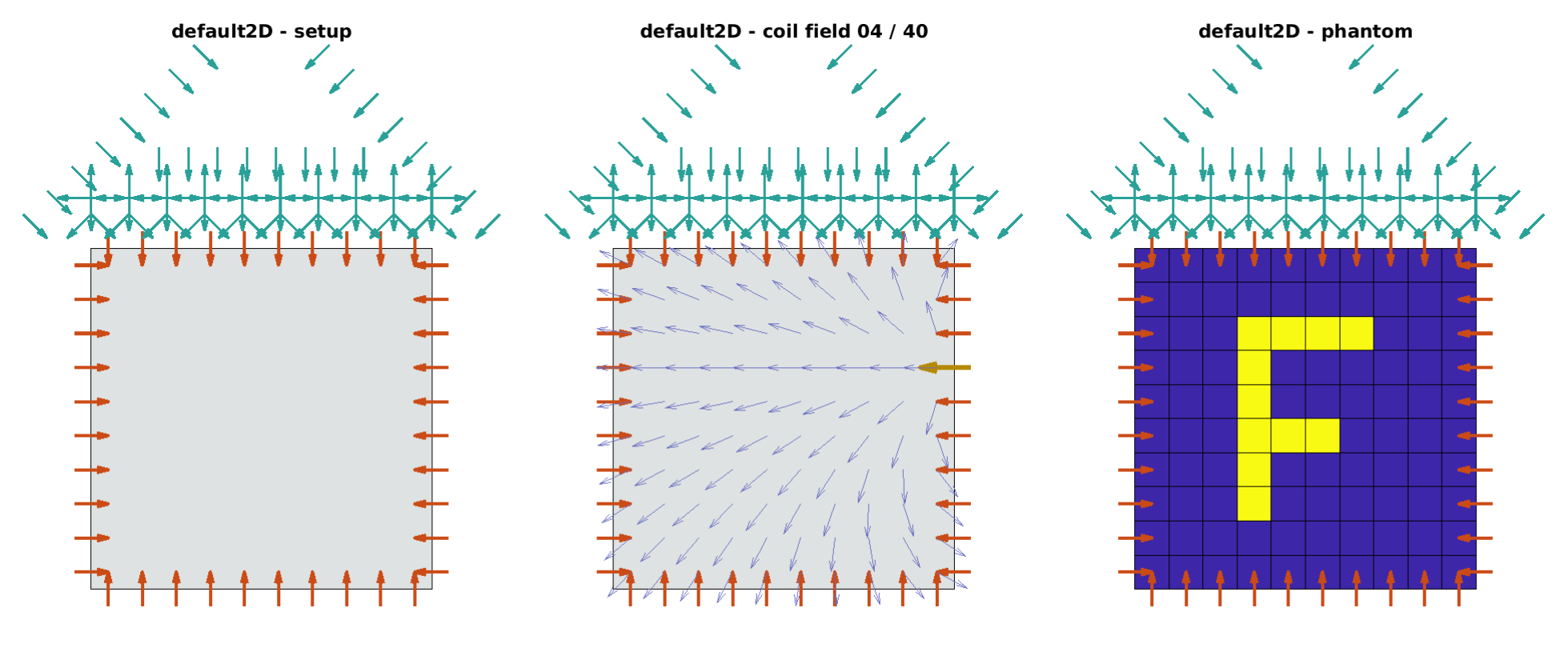}\\
		\includegraphics[width=\textwidth]{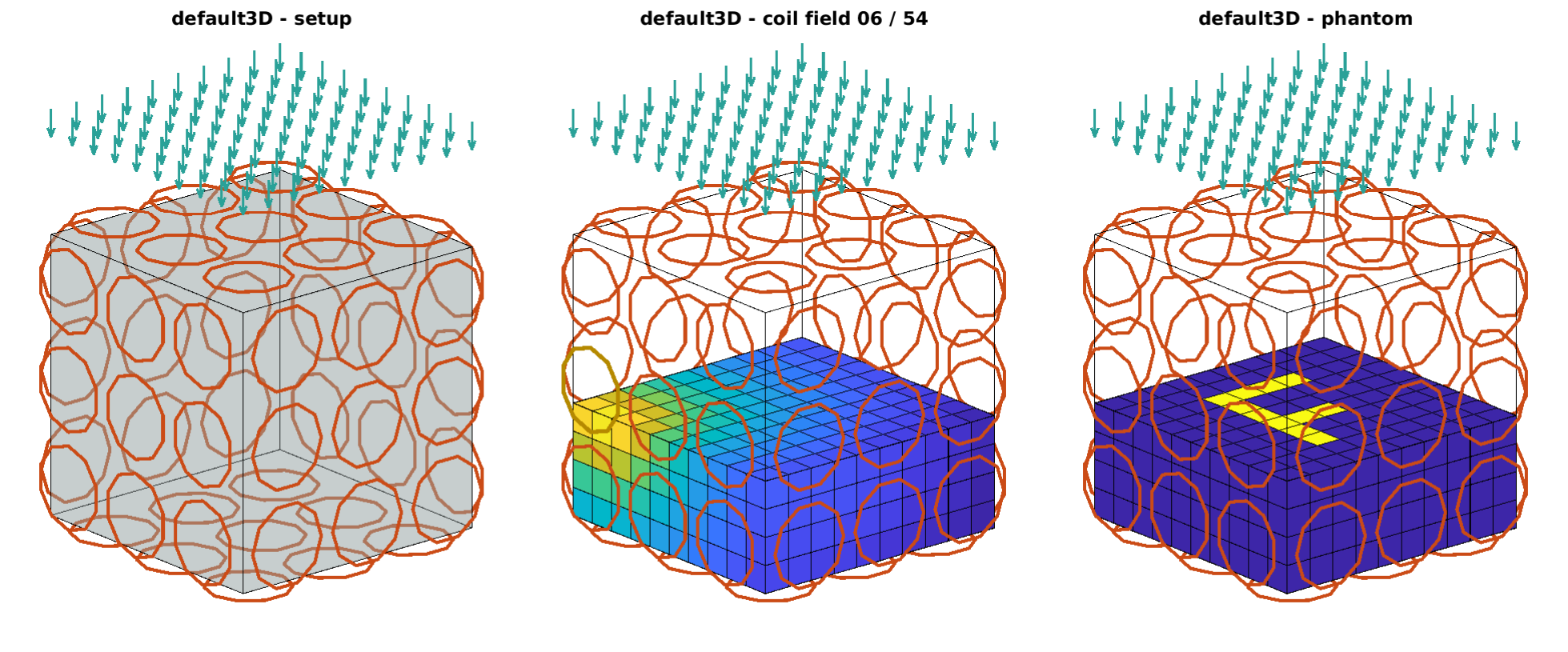}\\
		\includegraphics[width=\textwidth]{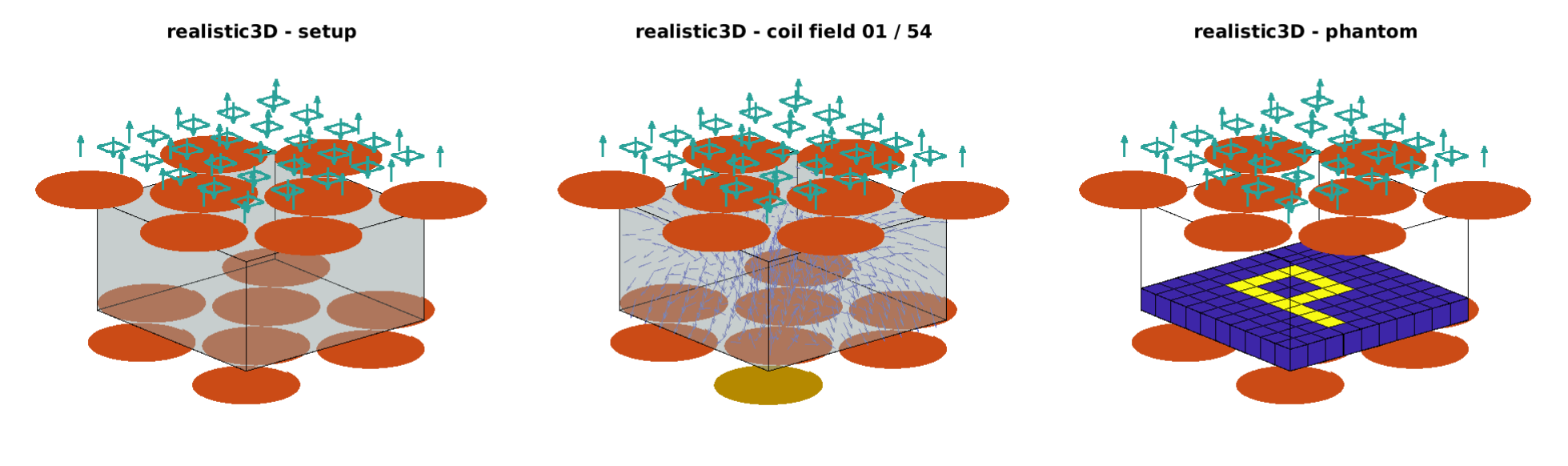}
	\end{mymarginbox}
	\caption{Visualization of the setups \emph{'default2D'} (top row), \emph{'default3D'} (middle row) and \emph{'realistic3D'} (bottom row). The left column shows the respective setup visualization. In the middle column one coil is active and the corresponding magnetic field is visualized. Here the setups \emph{default2D} and \emph{realistic3D} are visualizing the magnetic fields as arrows, whereas \emph{default3D} shows a volume visualization of the fields magnetic field strength in each voxel. The right column shows the setup with an embedded phantom. The \emph{default2D} and \emph{default3D} setups are using a simple \texttt{'F'} phantom in 2D and 3D respectively. The \emph{realistic3D} setup is using a \texttt{'P'} phantom, where the P is positioned in the bottom layer. Coils are colored red, sensors green, magnetic fields blue.}
	\label{im:vis}
\end{figure}
\simrx provides tools to visualize the region of interest (\texttt{drawROI.m}), coils (\texttt{drawCoils.m}), sensors (\texttt{drawSensors.m}), magnetic fields (\texttt{drawField.m}) and 3D volumes (\texttt{drawVolume.m}).

The function \texttt{visualizeMRX.m} is an wrapper that combines all of these functions into a single visualization tool.
It is equipped with multiple hotkeys that allows dynamic analysis of a given system.
With \texttt{visualizeMRX.m} we are able to create live previews of applied current patterns and visualize these fields with arrow fields or intensity maps.
Furthermore we can dynamically switch of each visualization module, e.g. to hide sensor arrays. 
The visualizations shown in \cref{im:vis} are all created using \texttt{visualizeMRX.m} only.
The figures can be recreated using the code provided in \mypath{./examples/visualization.m}.

%% file: sections/setups.tex
\section{Setups}\label{s:setups}
\begin{figure}
	\begin{mymarginbox}
		\begin{lstfloat}[H]
		\begin{verbatim}
		setups/                        % setups base folder
		┣━ default2D/                  % folder of the default2D setup
		┃  ┣━ README.m                 % additional information and script
		┃  ┣━ default.mrxsetup         % default variant setup file
		┃  ┣━ configs/                 % contains all configs for this setup
		┃  ┃  ┣━ all/                  % folder for configs that use all coils
		┃  ┃  ┃  ┣━ singleSequential.mrxcfg
		┃  ┃  ┃  ┃                     % config variant using sequential activation
		┃  ┃  ┃  ┣━ singleSequential2A.mrxcfg
		┃  ┃  ┃  :                     % variant with doubled coil current
		┃  ┃  ┃
		┃  ┃  ┣━ <config name>/        % another config for this setup
		┃  ┃  :  :
		┃  ┃
		┃  ┣━ raw/                     % raw data (via exportRawSetup.m)
		┃  ┗━ scripts/                 % scripts to create this config/setup
		┃     ┣━ createSetup.m         % create *.mrxsetup file
		┃     ┣━ createConfigs.m       % create *.mrxcfg in folder 'configs'
		┃     :
		┃
		┣━ <my setup>/                % my new setup
		┃  :
		┃
		:
		\end{verbatim}
		\caption[Setup folder structure]{Proposed setup folder structure, shown for included setup \emph{'default2D'}}
		\label{code:folderStructure}
		\end{lstfloat}
	\end{mymarginbox}
\end{figure}%
\begin{table}
	\begin{mymarginbox}
	\begin{subtable}{\textwidth}
		\begin{tabular}[t]{c||c|c|c|c|c|c|c|c|c}%
			&\texttt{x[m]}&\texttt{y[m]}&\texttt{z[m]}&\texttt{n$_x$}&\texttt{n$_y$}&\texttt{n$_z$}&\texttt{SensorID}&\texttt{ChannelID}&\texttt{GroupID}\\
			\hline Sensor 1&&&&&&&1&&\\
			\vdots&&&&&&&\vdots&&\\
			Sensor $N_s$&&&&&&&$N_s$&&\\
		\end{tabular}
		\caption{File: \texttt{sensors.dat}: sensor data structure}
		\label{tab:dataSensors}
	\end{subtable}\vspace{.5cm}
	\begin{subtable}{0.5\textwidth}
		\begin{tabular}[t]{c||c|c|c}%
			&\texttt{x[m]}&\texttt{y[m]}&\texttt{z[m]}\\
			\hline Coil 1&&&\\
			\vdots&&&\\
			Coil $N_c$&&&
		\end{tabular}
		\caption{File: \texttt{coilGrid.dat}:\newline coil position data structure}
		\label{tab:dataCoilGrid}
	\end{subtable}%
	\begin{subtable}{0.5\textwidth}
		\begin{tabular}[t]{c||c|c|c}%
			&\texttt{x[m]}&\texttt{y[m]}&\texttt{z[m]}\\
			\hline Segment 1&&&\\
			\vdots&&&\\
			Segment $N_l$&&&
		\end{tabular}
		\caption{File: \texttt{coilTemplate.dat}:\newline template coil data structure}
		\label{tab:dataCoilTemplate}
	\end{subtable}\vspace{.5cm}
	\begin{subtable}{0.5\textwidth}
		\begin{tabular}[t]{c||c|c|c}%
			&\texttt{x[m]}&\texttt{y[m]}&\texttt{z[m]}\\
			\hline Voxel 1&&&\\
			\vdots&&&\\
			Voxel $N_v$&&&
		\end{tabular}
		\caption{File: \texttt{voxelGrid.dat}:\newline voxel grid data structure}
		\label{tab:dataVoxelGrid}
	\end{subtable}%
	\begin{subtable}{0.5\linewidth}
		\begin{tabular}[t]{c||c}%
			&\texttt{current[A]}\\
			\hline Coil 1&\\
			\vdots&\\
			Coil $N_c$&
		\end{tabular}
		\caption{File: \texttt{dataset.01.currents.dat}:\newline coil currents data structure}
		\label{tab:dataCurrents}
	\end{subtable}\vspace{.5cm}
	\begin{subtable}{\linewidth}
		\begin{tabular}[t]{l||c|c|c|c|c}%
			&\texttt{$\Delta B$[fT]}&\texttt{SensorID}&\texttt{ChannelID}&\texttt{GroupID}&\texttt{CoilNo}\\
			\hline Sensor 1&&&&&1\\
			\vdots&&&&&\vdots\\
			Sensor $N_s$&&&&&1\\
			Sensor 1&&&&&$N_c$\\
			\vdots&&&&&\vdots\\
			Sensor $N_s$&&&&&$N_c$
		\end{tabular}
		\caption{File: \texttt{dataset.01.relax.dat}: sensor data table structure}
		\label{tab:dataRelax}
	\end{subtable}
	\end{mymarginbox}
	\caption{Overview of the data structures, that are used for text based datasets. \textbf{Note}: physical units are given in the brackets. The total number of sensors, coils, segments and voxel is given by $N_s$,$N_c$,$N_l$ and $N_v$ respectively.}
	\label{tab:dataStructure}
\end{table}%

This toolbox contains three setups, namely \emph{'default2D'}, \emph{'default3D'} and \emph{'realistic3D'}.
These can be used to test the functionalities of the \simrx toolbox.
We will first outline the used folder structure in \cref{ss:setupfolderstructure} and then give a short introduction to the included setups in \cref{ss:examplesetups}.

\textbf{Please note:} Due to internal use of the provided setups for short example codes, we firmly request to not edit the setups within the \simrx toolbox folder. Please create a new \emph{'setups'} folder with copies of the provided setups outside the toolbox and make changes in there. This keeps the \simrx directories clean and preserves its integrity for our internal example scripts.

\subsection{Folder structure}\label{ss:setupfolderstructure}
The following folder structure is proposed for \texttt{setup} and its respective \texttt{config} datasets (see \cref{code:folderStructure} as a reference and visual representation of the folder structure).

For each unique setup a base folder \mypath{./setups/<my setup>}is created and all related \texttt{configs}, scripts and datasets are included in this folder.
At this folders base level the \texttt{.mrxsetup} variant files are saved that describe the setup.

The folder \mypath{./setups/<setup>/configs} contains subfolders for each config that is compatible with the respective setup.
The \texttt{.mrxconfig} files for all available config variants are found in \mypath{./setups/<setup>/configs/<config name>/}.

The folder \mypath{./setups/<setup>/raw/} is reserved for raw data exports, that has been created by functions found in \mypath{./simulation/rawTools/} (see \cref{sss:rawExIm} for the raw export/import system).

Finally the folder \mypath{./setups/<setup>/scripts/} contains scripts for this specific setup.
This includes scripts for the creation of \texttt{.mrxsetup} and compatible \texttt{.mrxconfig} datasets.

\subsection{Setup and Config Variants}\label{ss:setupvariants}
As we have seen in \cref{sss:setup} a setup is essentially defined by the positioning of its entities.
In \cref{ss:coilDiscretization} however we shortly mentioned that a coil can be approximated by piecewise constant segments and likewise by a magnetic dipole.
Additionally the precision of the coil approximation has impact on the simulation speed of the system, since more segment fields have to be derived and combined.
This motives the concept of setup variants: The setups positioning of the entities is the same for each variant, but the physical properties of the entities can be varied.
An obvious example can be given by examining the \emph{default3D} setup: The \emph{default} setup variant is of this setup is shown in \cref{im:vis}.
Here the coil is discretized by $10$ straight conductor segment.
Obvious setup variants include the approximation of the coil with a dipole element, or an approximation with more coil segments.
Note: Since the 2D case is limited to dipole based coils, the concept of setup variants does only makes sense in 3D setups.

On the other hand, the concept of config variants can be applied for 2D and 3D setups alike.
We have discussed the structure of configs in \cref{sss:config} and can be essentially defined by its set of active coils and sensors, combined with the applied current and measurement patterns.
So more generally we can first sort configs by the set of active entities \texttt{coilsActive} and \texttt{sensorsActive}.
As such the example configs given in \cref{code:folderStructure} collected by the config name \emph{all}, are all using all available entities form the corresponding setup.
A config variant then describes the different current and measurement patterns.
The most default approach is the \emph{singleSubsequent} config variant, which relates to a subsequent activation of each coil one after another, where all responses are measured by all available sensors.

With this structure we can combine setup and config variants as desired.
For example this allows to easily exchange parts of the setup/config system without excessive reprogramming of the other parts.

\subsection{Example setups}\label{ss:examplesetups}
The respective folder \mypath{./setups/<setup>/scripts/} contain scripts for the creation of the following datasets.
Furthermore the file \texttt{README.m} in \mypath{./setups/<setup>} can be executed to run all creation scripts for that setup at once.

\subsubsection{Fully synthetic dataset: \emph{'default2D'} and \emph{'default3D'}}\label{subs:sytheticdataset}
In \simrx a 2D (\emph{'default2D'}) and a 3D (\emph{'default3D'}) example is available.
These can be created using the respective MATLAB scripts \texttt{createSetup.m} and \texttt{createConfigs.m} (the 2D scrips are shown in \cref{code:2Dsetup} and \cref{code:2Dconfig} as seen in \cref{sss:setup} and \cref{sss:config}).
A visualization of both datasets is available in \cref{ss:vis}.

\subsubsection{A 3D dataset from formatted text files: \emph{'realistic3D'}}\label{subs:loadext}
With \simrx it is possible to load datasets from text files, save in the \texttt{.mrxsetup} and \texttt{.mrxcfg} data structure, as well as pre-processing for simulation and reconstructions tasks.

In the subfolder \mypath{./setups/realistic3D/scripts} the script \texttt{createRawDataset.m} is used to create the following files in folder \mypath{./setups/realistic3D/rawData}:

\begin{mymarginbox}	
	\begin{tabular}{@{}L{\textwidth}@{}}%
	\texttt{sensors.dat}\\
	\texttt{coilGrid.dat}\\
	\texttt{coilTemplate.dat}\\
	\texttt{voxelGrid.dat}\\
	\texttt{dataset.01.currents.dat}\\
	\texttt{dataset.01.relax.dat}
	\end{tabular}
\end{mymarginbox}

The file structure of the created data is as follows.

\paragraph{Sensor Information (\Cref{tab:dataSensors}, File \texttt{sensors.dat}):}
The table (see \cref{tab:dataSensors}) that is used in file \texttt{sensors.dat}) stores all necessary sensor information.
Each row defines a sensor unit with properties defined by the columns as follows. Columns $1-3$ define the \texttt{x}, \texttt{y} and \texttt{z} position as a translation vector. Columns $4-6$ define the orientation of the sensors measure direction. Columns $7-9$ are used to identify the given coil: column $7$ holds an unique sensorID, column $8$ holds information about the used data channel and column $9$ is used to defined sensor groups.

\paragraph{Coil Information (\Cref{tab:dataCoilGrid}, File \texttt{coilGrid.dat} and \cref{tab:dataCoilTemplate}, File \texttt{coilTemplate.dat}):}
The file \texttt{coilGrid.dat} contains positioning information for every activation coil in the system (see \cref{tab:dataCoilGrid}).
Each row defines a coil position with columns $1-3$ defining the translation vector for direction \texttt{x}, \texttt{y} and \texttt{z}.
As described in \cref{ss:coilDiscretization} and \cref{sss:setup}, the coil is a composition of multiple conductor segments.
The coils single segments are stored as a list of points in file \texttt{coilTemplate.dat} (see \cref{tab:dataCoilTemplate}).

\paragraph{Voxel Grid Information (\Cref{tab:dataVoxelGrid}, File: \texttt{voxelGrid.dat}):}
The file \texttt{voxelGrid.dat} contains a list of used voxels in the current setup (see \cref{tab:dataVoxelGrid}).
Again columns $1-3$ define the positions of the voxel midpoints in \texttt{x}, \texttt{y} and \texttt{z}.

\paragraph{Current Information (\Cref{tab:dataCurrents}, File: \texttt{dataset.01.currents.dat}):}
The \cref{tab:dataCurrents} is used as data scheme for file \texttt{dataset.01.currents.dat} and contains a list of currents in Ampere $\left[\si{A}\right]$ that are applied to the coils.
Currently, for external data, only subsequent coil patterns are supported.
This means that the number of givencurrents has to fit the number of coils as of \texttt{coilGrid.dat}.

\paragraph{Measurements Information (\Cref{tab:dataRelax}, File: \texttt{dataset.01.relax.dat}):}
The data structure defined in \cref{tab:dataRelax}) is used in file \texttt{dataset.01.relax.dat}.
The first column contains the change in the magnetic response $\Delta B \left[\si{fT}\right]$ after an coil activation (compare with equation \eqref{eq:measmagnetization}).
Column $2-5$ provide information about the sensorID, channelID, groupID and the coil/current pattern that was used for that measurement.

The raw dataset then found in \mypath{./setups/realistic3D/rawData} can be loaded and parsed with the scripts provided in \mypath{./setups/realistic3D/scripts}, namely \texttt{createSetup.m} and finally \texttt{createConfigs.m}.
As expected, this creates a \texttt{.mrxsetup} and \texttt{.mrxcfg} file.
A visualization of this dataset is available in \cref{ss:vis}.
With \texttt{loadDataset.m} the data can be loaded and cleaned up in regards of faulty or unused sensors.
Finally, with \texttt{simulateMeasurement.m} a simulated measurement is created based on the given dataset.
Here the magnetic susceptibility $\chi$ and amount of the particles (in $\left[\si{mg}\right]$) are required to define the phantom properly.

As for the other examples, a script \texttt{README.m} is available, that can be used to execute the individual scripts in the recommended order.

%% file: sections/examples.tex
\section{Examples}\label{s:examples}
The folder \mypath{./exmaples} contains scripts to illiustrate the workflow that is required to handle simulated and real experimenal data with the \simrx toolbox.
It make use of the provided setups presented in \cref{ss:examplesetups}.

\begin{enumerate}
\item \texttt{ExampleA.m} loads the \emph{'default2D'} setup and a eligible config file (see \cref{subs:sytheticdataset}) and simulates the corresponding system matrix.
\item \texttt{ExampleB.m} loads the \emph{'default3D'} setup and a eligible config file (see \cref{subs:sytheticdataset}) and stores the simulated system in the raw data format (see \cref{sss:rawExIm}).
Finally the raw data is imported and combined to a proper system matrix.
\item \texttt{ExampleC.m} loads the \emph{'realistic3D'} setup and a eligible config file (see \cref{subs:loadext}).
Then measured data are loaded, defective sensors are removed and the measured data is compared to a simulated measurement.
With this example, data from real experiment can be processed.
\end{enumerate}

%% file: sections/remarks.tex
\section{Remarks}
This simulation toolbox does not guarantee its correctness and closeness to reality.
\simrx is not qualified for commercial usage.

%% file: sections/documentation.tex
\newpage\section{Documentation}
In the following we compiled a list of all functions that come with the \simrx toolbox.
Here we only give an short explanation what this function is supposed to do.
For usage instructions, arguments and optional parameters, please read the documentation in the head of each file.

\subsection{File List}
\begin{tabularx}{\linewidth}{l||X}%
	\multicolumn{2}{l}{\ }\\\multicolumn{2}{l}{\mypath{./}}\\\hline
	\texttt{initSiMRX.m}&adds toolbox subfolder to matlab path\\
	\multicolumn{2}{l}{\ }\\\multicolumn{2}{l}{\ }\\\multicolumn{2}{l}{\mypath{./configuration/}}\\\hline
	\texttt{checkCompatibility.m}&checks if a given setup is compatible with the given config.\\
	\texttt{isConfigValid.m}&checks if a given config is valid\\
	\texttt{isSetupValid.m}&checks if for a given setup is valid\\
	\texttt{loadConfig.m}&loads a mrx setup from a file\\
	\texttt{loadExampleMRXDataset.m}&loads a 2D or 3D example dataset\\
	\texttt{loadSetup.m}&loads a mrx setup from a file\\
	\texttt{saveConfig.m}&saves a mrx config to a config file\\
	\texttt{saveSetup.m}&saves a mrx setup to a setup file\\
	\multicolumn{2}{l}{\ }\\\multicolumn{2}{l}{\ }\\\multicolumn{2}{l}{\mypath{./configuration/buildTools/}}\\\hline
	\texttt{calibrateCoil.m}&derives scaling factor to calibrate idealized coil\\
	\texttt{cleanupPatterns.m}&removes empty patterns from  provided current and measurement patterns\\
	\texttt{createCoilLoop.m}&creates a 3D point cloud that describes a coil\\
	\texttt{createEntityArray.m}&creates an array of entities\\
	\texttt{createPattern.m}&creates a current or measurement pattern\\
	\texttt{getROI.m}&returns the smallest cartasian intervals that contain the point cloud defined by a given voxel grid\\
	\texttt{parseCoils.m}&uses a template point cloud and place in multiple places\\
	\texttt{relocateStructure.m}&moves a given 3D structure to a new position and orientation\\
	\texttt{setCoilLoopFactor.m}&rescales a coils normal vector by a factor\\
	\multicolumn{2}{l}{\ }\\\multicolumn{2}{l}{\ }\\\multicolumn{2}{l}{\mypath{./examples/}}\\\hline
	\texttt{exampleA.m}&runs the default2D setup illustrating usage of setup and config files\\
	\texttt{exampleB.m}&runs the default3D setup illustrating usage of raw files\\
	\texttt{exampleC.m}&runs the realistic3D setup illustrating usage of external data structures\\
	\texttt{README.m}&prepares all setups provided in the toolbox\\
	\texttt{visualization.m}&show capabilities of visualizeMRX\\
	\multicolumn{2}{l}{\ }\\\multicolumn{2}{l}{\ }\\\multicolumn{2}{l}{\mypath{./misc/}}\\\hline
	\texttt{isOctave.m}&returns true if executed in octave\\
	\texttt{loadBinary.m}&loads a dataset in the provided path\\
	\texttt{parsave.m}&saves a variable even within a parfor loop\\
	\texttt{parsaveBinary.m}&stores the dataset data in path\\
	\texttt{rawDataLoader.m}&data loader for MRX data\\
\end{tabularx}\newpage
\begin{tabularx}{\linewidth}{l||X}%
	\multicolumn{2}{l}{\mypath{./misc/io/}}\\\hline
	\texttt{decoratedBox.m}&prints a text and surrounds it with a visual box\\
	\texttt{fpo.m}&prints given string\\
	\texttt{fpon.m}&prints given string + newline\\
	\texttt{fpop.m}&prints pair of strings\\
	\texttt{logger.m}&handles printouts and logs with verbose level\\
	\multicolumn{2}{l}{\ }\\\multicolumn{2}{l}{\ }\\\multicolumn{2}{l}{\mypath{./phantom/}}\\\hline
	\texttt{createPhantom.m}&creates commonly used phantoms for MRX experiments\\
	\texttt{phantom3.m}&creates analytical 3D phantoms\\
	\multicolumn{2}{l}{\ }\\\multicolumn{2}{l}{\ }\\\multicolumn{2}{l}{\mypath{./setups/}}\\\hline
	\texttt{README.m}&prepares all setups provided in the toolbox\\
	\multicolumn{2}{l}{\ }\\\multicolumn{2}{l}{\ }\\\multicolumn{2}{l}{\mypath{./setups/default2D/scripts/}}\\\hline
	\texttt{createConfigs.m}&creates a config for the default2D setup\\
	\texttt{createSetup.m}&creates setup called default2D\\
	\multicolumn{2}{l}{\ }\\\multicolumn{2}{l}{\ }\\\multicolumn{2}{l}{\mypath{./setups/default3D/scripts/}}\\\hline
	\texttt{createConfigs.m}&creates a config for the default3D setup\\
	\texttt{createSetup.m}&creates setup called default3D\\
	\multicolumn{2}{l}{\ }\\\multicolumn{2}{l}{\ }\\\multicolumn{2}{l}{\mypath{./setups/realistic3D/scripts/}}\\\hline
	\texttt{createConfigs.m}&creates a config for the realistic3D setup\\
	\texttt{createRawData.m}&creates setup and config raw files as plain text files\\
	\texttt{createSetup.m}&creates setup called realistic3D\\
	\texttt{loadMeasuredData.m}&loads a measurement from a raw file\\
	\texttt{simulateMeasuredData.m}&simulates a measurement and compares it with reference measurement\\
	\multicolumn{2}{l}{\ }\\\multicolumn{2}{l}{\ }\\\multicolumn{2}{l}{\mypath{./simulation/}}\\\hline
	\texttt{createMRXMatrix.m}&simulates MRX system based on a given setup and config\\
	\multicolumn{2}{l}{\ }\\\multicolumn{2}{l}{\ }\\\multicolumn{2}{l}{\mypath{./simulation/rawTools/}}\\\hline
	\texttt{isRawValid.m}&checks if raw folder contains all necessary information\\
	\texttt{rawDeriveSetup.m}&derives raw files\\
	\texttt{rawExportSetup.m}&simulates and exports a setup in raw format\\
	\texttt{rawExportSetupPar.m}&simulates and exports a setup in raw format\\
	\texttt{rawImportMagneticFields.m}&imports magnetic field data from raw files\\
	\texttt{rawImportMatrix.m}&imports system matrix from raw files\\
	\texttt{rawImportMeasurement.m}&imports measurement from raw files\\
	\texttt{rawImportSetup.m}&imports a raw dataset from path\\
\end{tabularx}\newpage
\begin{tabularx}{\linewidth}{l||X}%
	\multicolumn{2}{l}{\mypath{./simulation/simTools}}\\\hline
	\texttt{applyPatterns.m}&applies a coil activation scheme to a simulated raw system\\
	\texttt{createCoilField.m}&derives a magnetic field\\
	\texttt{createDipoleField.m}&derives a magnetic field\\
	\texttt{createExcitationFields.m}&calculates magnetic fields induced by coils or dipoles\\
	\texttt{createMRXMatrixByCoil.m}&derives a MRX system matrix\\
	\texttt{createMRXMatrixByPattern.m}&derives a MRX system matrix\\
	\texttt{createRelaxationFields.m}&derives magnetic fields in sensors\\
	\texttt{createVoxelGrid.m}&creates a voxel grid for a given region of interest\\
	\texttt{deriveMeasurement.m}&derives a measurement\\
	\multicolumn{2}{l}{\ }\\\multicolumn{2}{l}{\ }\\\multicolumn{2}{l}{\mypath{./visualization/}}\\\hline
	\texttt{visualizeMRX.m}&visualization tool for MRX setups\\
	\multicolumn{2}{l}{\ }\\\multicolumn{2}{l}{\ }\\\multicolumn{2}{l}{\mypath{./visualization/drawTools/}}\\\hline
	\texttt{drawCoils.m}&visualizes coils\\
	\texttt{drawField.m}&visualizes magnetic fields\\
	\texttt{drawPattern.m}&visualizes a magnetic field induced by a coil pattern\\
	\texttt{drawROI.m}&visualizes the region of interest\\
	\texttt{drawSensors.m}&visualizes sensors\\
	\texttt{drawVolume.m}&visualizes volume data\\
	\multicolumn{2}{l}{\ }\\\multicolumn{2}{l}{\ }\\\multicolumn{2}{l}{\mypath{./visualization/misc/}}\\\hline
	\texttt{flipXY.m}&mirrors in Y direction and swaps X and Y coordinates\\
	\texttt{getSetupBounds.m}&derives the effective region of interest that contains all setups entities\\
	\texttt{loadVisPresets.m}&presets for visualization tools\\
\end{tabularx}